\documentclass{article}
\usepackage{spconf,amsmath,graphicx}
\usepackage{amssymb}
\usepackage{url}

\title{TRISTouNET: TRIPLET LOSS FOR SPEAKER TURN EMBEDDING}

\name{Herv\'{e} Bredin}
\address{LIMSI, CNRS, Universit\'{e} Paris-Saclay}

\begin{document}
%
\maketitle
\begin{abstract}
\emph{TristouNet} is a neural network architecture based on Long Short-Term Memory recurrent networks, meant to project speech sequences into a fixed-dimensional euclidean space. Thanks to the triplet loss paradigm used for training, the resulting sequence embeddings can be compared directly with the euclidean distance, for speaker comparison purposes. Experiments on short (between 500ms and 5s) speech turn comparison and speaker change detection show that \emph{TristouNet} brings significant improvements over the current state-of-the-art techniques for both tasks.

\end{abstract}
\section{Introduction}
\label{sec:intro}

Given a speech sequence~$x$ and a claimed identity~$a$, speaker verification aims at accepting or rejecting the identity claim. It is a supervised binary classification task usually addressed by comparing the test speech sequence~$x$ to the enrollement sequence~$x_a$ uttered by the speaker~$a$ whose identity is claimed.
Speaker identification is the task of determining which speaker (from a predefined set of speakers $a \in \mathcal{S}$) has uttered the sequence~$x$. It is a supervised multiclass classification task addressed by looking for the enrollement sequence~$x_a$ the most similar to the test speech sequence~$x$.
Speaker diarization is the task of partitioning an audio stream into homogeneous temporal segments according to the identity of the speaker. It is broadly addressed as the series of three steps: speech activity detection, speaker change detection (\emph{i.e.} finding boundaries between any two different speakers), and speech turn clustering.

Whether we address speaker verification, speaker identification, or speaker diarization, it all boils down to finding the best pair ($f$, $d$) of representation function $f$ and comparison function $d$ with the following ideal property. Given a speech sequence~$x_a$ uttered by a given speaker, any speech sequence~$x_p$ uttered by the same speaker should be \emph{closer} to $x_a$ than any speech sequence~$x_n$ uttered by a different one:
\begin{equation}
  d(f(x_a), f(x_p)) < d(f(x_a), f(x_n))
  \label{eq:property}
\end{equation}

Judging from the organization of the \emph{NIST i-vector Machine Learning Challenge}~\cite{Greenberg2014}, the i-vector approach~\cite{Dehak2011} has become the \emph{de facto standard} for $f$ as far as speaker recognition is concerned.
Hence, given a common i-vector implementation, the objective for participants to this challenge is to design the best comparison function $d$.
In this paper, we address the dual problem: choosing $d$ as the euclidean distance, we want to find a representation function $f$ that has the property described in Equation~\ref{eq:property}. Practically, based on a carefully designed loss function, we propose to train a speech sequence embedding based on recurrent neural networks to get closer to this optimal function $f$.

The i-vector approach has also become the state-of-the-art for speaker diarization~\cite{Dupuy2014}.
However, due to its sensitivity to sequence duration~\cite{Sarkar2012}, it is only used once short speech turns have  been clustered into larger groups using Bayesian Information Criterion (BIC)~\cite{Chen1998} or Gaussian divergence~\cite{Barras2006}. These two techniques are still commonly used for short (\emph{i.e.} shorter than 5 seconds) speech turn segmentation and clustering. In this paper, we show that the proposed embedding outperforms both approaches and leads to better speaker change detection results.


\section{Related work}
\label{sec:prior}

Using the triplet loss~\cite{Wang2014,Hoffer2015} to train euclidean embeddings has been recently and successfully applied to face recognition and clustering in~\cite{Schroff2015}. We use the triplet loss and triplet sampling strategy they proposed. Going with the euclidean distance and unitary embeddings was also inspired by it. The main difference lies in the choice of the neural network architecture used for the embedding.
While convolutional neural networks are particularly adapted to (multi-dimensional) image processing and were used in~\cite{Schroff2015}, we went with recurrent neural networks (more precisely, bi-directional long short-term memory networks, BiLSTM) that are particularly adapted to sequence modeling~\cite{Hochreiter1997} and were first used for speech processing in~\cite{Graves2005}.

The idea of using deep neural networks to learn a representation function adapted to speaker recognition is not novel~\cite{Konig1998,Yella2014}.
Back in 1998, \emph{Konig et al.} trained a multilayer perceptron (MLP) where the input consists of cepstral coefficients extracted from a sequence of frames, and the output layer has one output per speaker in the training set~\cite{Konig1998}.
The activations of (bottleneck) hidden layers are then used as the representation function for later speaker recognition experiments using (back then, state-of-the-art) Gaussian Mixture Models (GMM).
The main limitations for this kind of approaches is summarized nicely by \emph{Yella et al.} in their recent paper~\cite{Yella2014}:  \emph{``we \underline{hypothesize} that the hidden layers of a network trained in this fashion \underline{should} transform spectral features into a space more conducive to speaker discrimination.''}.
In other words, we are not quite sure of the efficiency of this internal representation as it is not the one being optimized during training -- these approaches still require a carefully designed comparison function $d$ (based on GMMs for~\cite{Konig1998} or Hidden Markov Models for~\cite{Yella2014}). Our approach is different in that the representation function $f$ is the one being optimized with respect to the fixed euclidean distance $d$.

That being said, \cite{Yella2014} is very similar to our work in that their neural network is given pairs of sequences as input, and is trained using binary cross-entropy loss to decide whether the two sequences are from the same speaker or from two different speakers.
With pairs of 500ms speech sequences, they report a 35\% error rate on this task.
As depicted in Figure~\ref{fig:triplet}, the main difference with our approach lies in the fact the we use triplets of sequences (instead of pairs) and optimize the shared embedding directly thanks to the triplet loss (in place of the intermediate binary cross-entropy loss).

Recently, LSTMs have been particularly successful for automatic speech recognition~\cite{Graves2013}.
They have also been applied recently to speaker adaptation for acoustic modelling~\cite{Miao2015,Tan2016}.
However, to the best of our knowledge, it is the first time they are used for an actual speaker comparison task, and \emph{a fortiori} for speaker turn embedding.

\section{Triplet loss for sequence embedding}
\label{sec:embedding}

\begin{figure}[htb]
  \centering
  \includegraphics[width=0.7\linewidth]{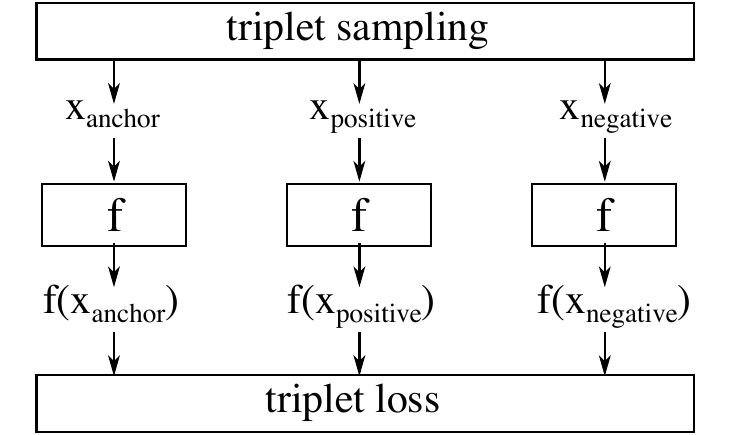}
  \caption{Triplet loss training}
  \label{fig:triplet}
\end{figure}

Figure~\ref{fig:triplet} summarizes the main idea behind triplet loss embeddings.
During training, the \emph{triplet sampling} module generates $(x_a, x_p, x_n)$ triplets where $x_a$ are features extracted from a sequence (called \emph{anchor}) of a given speaker, $x_p$ are from another sequence (called \emph{positive}) from the same speaker, and $x_n$ are from a sequence (called \emph{negative}) from a different speaker.
Then, all three feature vectors (or sequences of vectors, in our case) are passed through the neural network embedding $f$. Finally, the \emph{triplet loss}~\cite{Schroff2015} minimizes the distance between the embeddings of the \emph{anchor} and \emph{positive}, and maximizing the distance between the \emph{anchor} and \emph{negative}.

\subsection{Triplet loss}
\label{ssec:triplet_loss}

Let $\mathcal{T}$ be the set of all possible triplets $\tau = (x_a^\tau, x_p^\tau, x_n^\tau)$ in the training set.
The triplet loss is motivated by the Equation~\ref{eq:property} introduced earlier, and tries to achieve an even better separation between positive and negative pairs by adding a \emph{safety} margin $\alpha \in \mathbb{R}^+$. For any triplet $\tau$, we want $\Delta_{\tau} + \alpha < 0$ where
\begin{equation}
  \Delta_{\tau} = \|f(x_a^\tau) - f(x_p^\tau)\|_2^2 - \|(f(x_a^\tau) - f(x_n^\tau)\|_2^2
\end{equation}

\noindent More precisely, the loss that we try to minimize is defined as
\begin{equation}
  \mathcal{L}(\mathcal{T}) = \sum_{\tau \in \mathcal{T}} \max(0, \Delta_{\tau} + \alpha)
  \label{eq:triplet_loss}
\end{equation}

\subsection{Triplet sampling strategy}
\label{ssec:triplet_sampling}

As discussed thoroughly in~\cite{Schroff2015}, it is not efficient nor effective to generate all possible triplets. Instead, one should focus on triplets that violate the constraint $\Delta_{\tau} + \alpha < 0$.
Any other triplet would not contribute to the loss and would only make training slower. Though we do plan to test other triplet sampling strategies in the future, we chose to go with the one called \emph{``hard negative''} in~\cite{Schroff2015}.

More precisely, after each epoch, we repeat the following sampling process.
First, we start by randomly sampling $n$~sequences from each of the $N$~speakers of the training set. This leads to a total of $Nn(n-1)/2$ \emph{anchor}-\emph{positive} pairs. Then, for each of those pairs, we randomly choose one \emph{negative} out of all $(N-1)n$ negative candidates, such that the resulting triplet~$\tau$ has the following properties: $\Delta_{\tau} + \alpha > 0$.

\subsection{\emph{TristouNet} sequence embedding}
\label{ssec:embedding_architecture}

\begin{figure}[htb]
  \centering
  \includegraphics[width=0.7\linewidth]{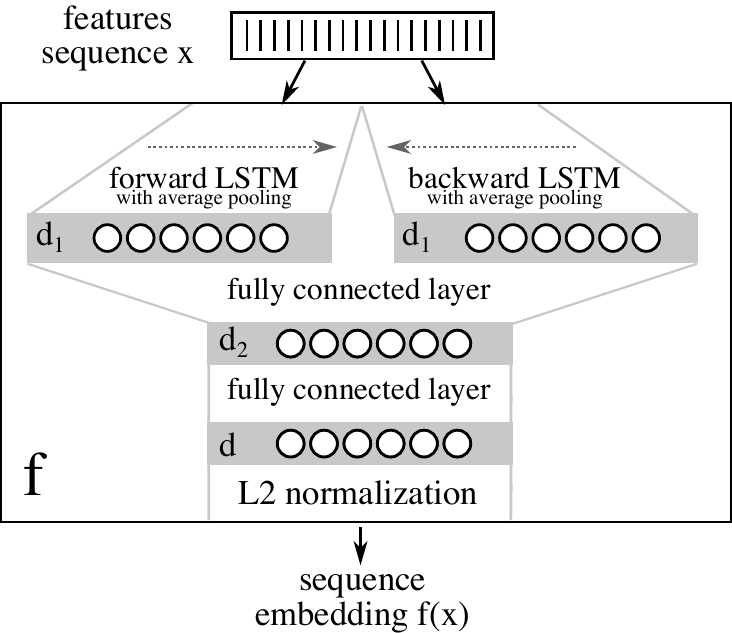}
  \caption{\emph{TristouNet} architecture}
  \label{fig:architecture}
\end{figure}

Figure~\ref{fig:architecture} depicts the topology of \emph{TristouNet}\footnote{\underline{tri}plet loss for \underline{s}peaker \underline{tu}rn neural \underline{net}work (colloquial French for \emph{gloomy}) }, the neural network $f$ we propose for sequence embedding.
Two Long Short-Term Memory (LSTM) recurrent networks~\cite{Hochreiter1997} (with $d_1$ units each) both take the feature sequence $x$ as input.
The first LSTM processes the sequence in chronological order, while the second goes backward.
Average pooling is applied to their respective sequence of outputs.
This leads to two $d_1$-dimensional output vectors which are then concatenated into one $2.d_1$-dimensional vector.
Returning only the average output has one advantage: projecting variable-length input sequences into a fixed-dimension space. However, in this paper, we only used fixed-length input sequences in order to evaluate how well the approach performs depending on the duration.
Two fully connected layers (with $d_2$ and $d$ units respectively) are then stacked.
The final output is $L^2$-normalized, constraining the final embedding to live on the $d$-dimensional unit hypersphere.

\section{Experiments}
\label{sec:experiments}

\subsection{Dataset}
\label{ssec:dataset}

The ETAPE TV subset contains 29 hours of TV broadcast (18h for training, 5.5h for development and 5.5h for test) from three French TV channels with news, debates, and entertainment~\cite{Gravier2012}.
Fine \emph{``who speaks when''} annotations were obtained on a subset of the training and development set using the following two-steps process: automatic forced alignement of the manual speech transcription followed by manual boundaries adjustment by trained phoneticians.
Overall, this leads to a training set of 13.8h containing $N=184$ different speakers, and a development set of 4.2h containing 61 speakers (out of which 18 are also in the training set). Due to coarser annotations, the test set is not used in this paper.

\subsection{Implementation details}

\noindent\textbf{Feature extraction.}
35-dimensional acoustic features are extracted every 20ms on a 32ms window using Yaafe toolkit~\cite{Mathieu2010}: 11 Mel-Frequency Cepstral Coefficients (MFCC), their first and second derivatives, and the first and second derivatives of the energy.
Both BIC and Gaussian divergence baselines rely on the same set of features (without derivatives, because it leads to better performance).

\noindent\textbf{Training.}
We use Keras~\cite{Chollet2016} deep learning library for training \emph{TristouNet}. The number of outputs is set to 16 for every layer (\emph{i.e.} $d_1 = d_2 = d = 16$). In particular, $d = 16$ means that the sequence embeddings live on the 16-dimensional unit hypersphere. We use $tanh$ activation function for every layer as well.
Every model (one for each sequence duration 500ms, 1s, 2s and 5s) is trained for 50 epochs, using margin $\alpha = 0.2$ as proposed in the original paper~\cite{Schroff2015}, and the RMSProp optimizer~\cite{Tieleman2012} with $10^{-3}$ learning rate.
Finally, the triplet sampling uses $n = 40$ random sequences per speaker, for a total of 143520 triplets per epoch.

\noindent\textbf{Reproducible research.} {\footnotesize\url{github.com/hbredin/TristouNet}} provides Python code to reproduce the experiments.

\subsection{\emph{``same/different''} toy experiment}
This first set of experiments aims at evaluating the intrinsic quality of the learned embedding.

\noindent\textbf{Protocol.}
100 sequences are extracted randomly for each of the 61 speakers in the ETAPE development set. The \emph{``same/different''} experiment consists in a binary classification task: given any two of those sequences, decide whether they were uttered by the same speaker, or two different speakers. This is achieved by thresholding the computed distance between sequences. We compare several approaches: Gaussian divergence~\cite{Barras2006}, Bayesian Information Criterion~\cite{Chen1998}, and the proposed embedding with euclidean distance.

\noindent\textbf{Evaluation metric.}
Two types of errors exist: a \emph{false positive} is triggered when two sequences from two different speakers are incorrectly classified as uttered by the same speaker, and a \emph{false negative} is when two sequences from the same speaker are classified as uttered by two different speakers. The higher (resp. lower) the decision threshold is, the higher the false negative (resp. positive) rate is (FNR, FPR). We report the equal error rate (EER), \emph{i.e.} the value of FPR and FNR when they are equal.

\begin{figure*}[htb]
  \centering
  \includegraphics[width=1.\linewidth]{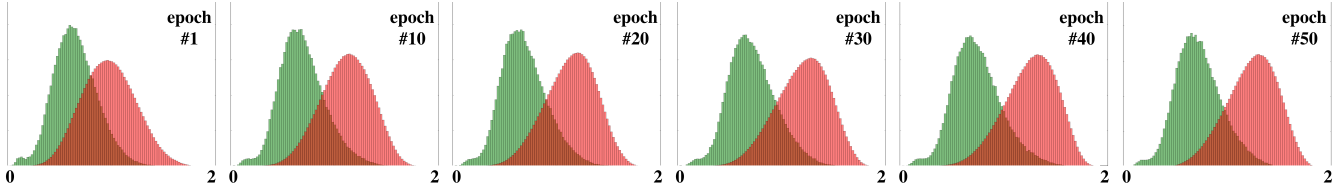}
  \caption{Distribution of distances between pairs of \emph{same speaker} (green) and \emph{different speaker} (red) 2s-sequence embeddings. This is based on every combinations of 100 random sequences of all speakers of the development set, every 10 training epochs.}
  \label{fig:distributions}
\end{figure*}

\begin{figure}[htb]
  \centering
  \includegraphics[width=1.\linewidth]{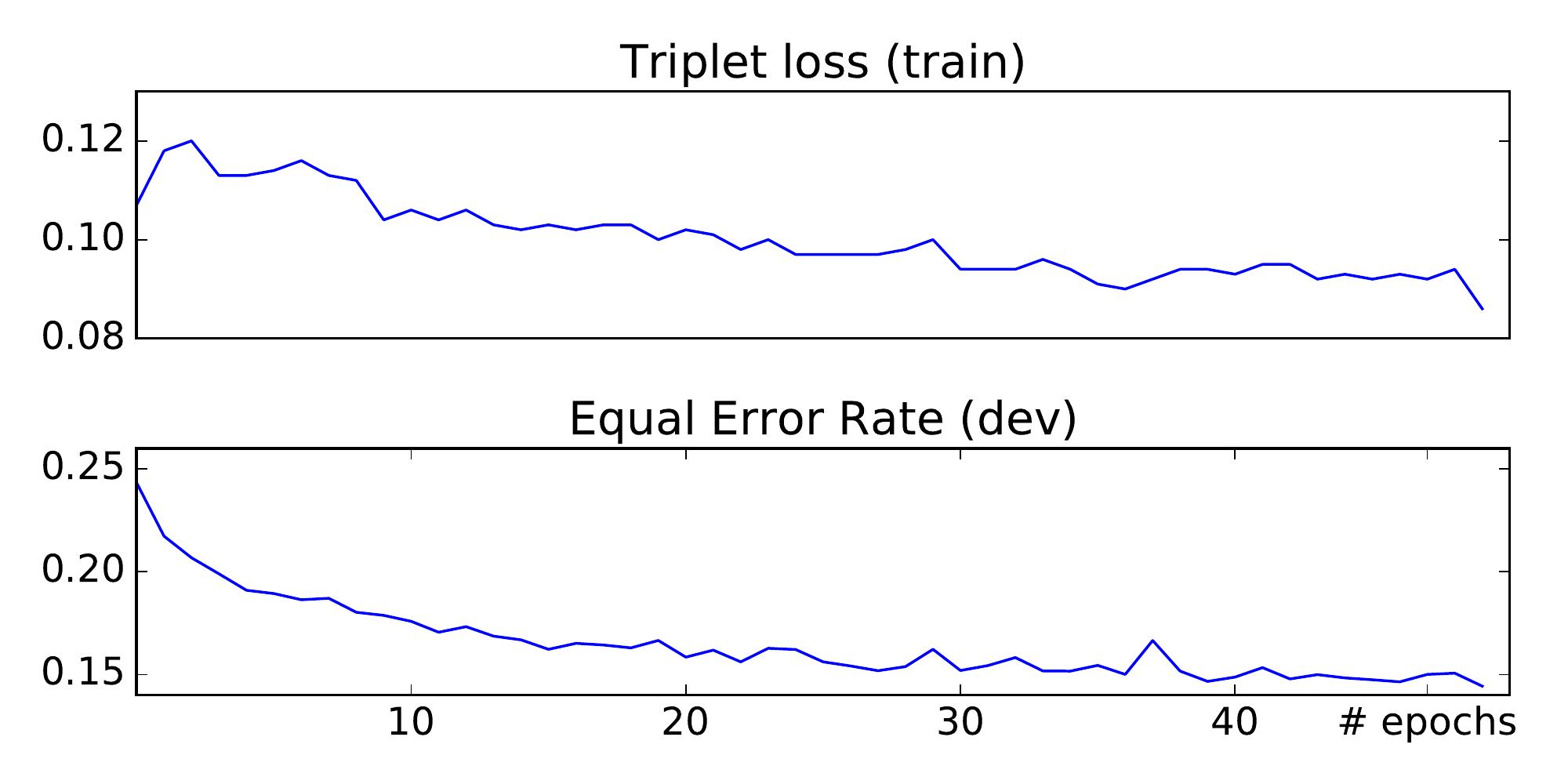}
  \caption{Top: triplet loss on the training set for the 2s sequence embedding after each epoch. Bottom: equal error rate obtained on the \emph{``same/different''} experiment on the development set after each epoch.}
  \label{fig:loss_vs_eer}
\end{figure}

\noindent\textbf{Training.}
Figures~\ref{fig:distributions}~and~\ref{fig:loss_vs_eer} illustrate how the intrinsic quality of the embedding (of 2s sequences) improves over time, during training. Figure~\ref{fig:distributions} clearly shows how the discriminative power of the embedding improves every 10 epochs: \emph{same} and \emph{different} speaker(s) distance distributions are progressively separating until convergence and no further significant improvement is observed.

\begin{figure}[htb]
  \centering
  \includegraphics[width=1.\linewidth]{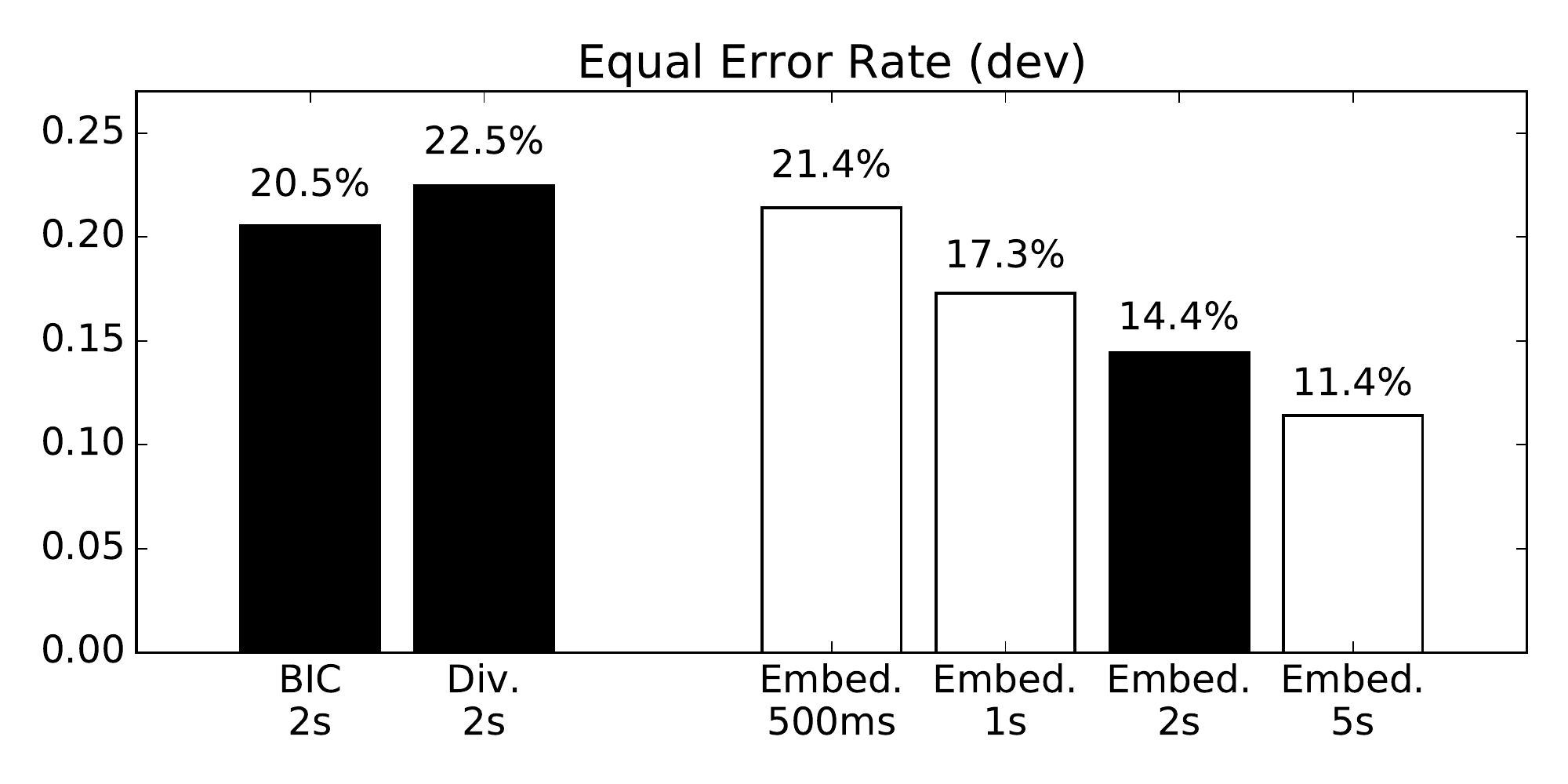}
  \caption{Results of the \emph{``same/different''} experiment for varying sequence duration. Embeddings are obtained after 50 epochs.}
  \label{fig:eer}
\end{figure}

\noindent\textbf{Results.}
Figure~\ref{fig:eer} summarizes the results. As expected, embeddings of longer sequences get better performance: EER decreases from $21.4\%$ for 500ms sequences down to $11.4\%$ for 5s sequences. Most importantly, our approach significantly outperforms the commonly used approaches (BIC and Gaussian divergence), bringing an absolute $6.1\%$ (or relative $30\%$) EER decrease for 2s sequence comparison. Note how the 500ms embedding is almost as good as the (four times longer) 2s BIC baseline approach.

\subsection{Speaker change detection}
\label{ssec:speaker_change_detection}

Speaker change detection consists in finding the boundaries between speech turns of two different speakers. It is often used as a first step before speech turns clustering in speaker diarization approaches.

\noindent\textbf{Protocol.}
For each files in the ETAPE development set, we compute the distance between two (left and right) 2s sliding windows, every 100ms. Peak detection is then applied to the resulting 1-dimensional signal by looking for local maxima within 1s context. A final thresholding step removes small peaks and only keeps large ones as speaker changes.

\noindent\textbf{Evaluation metric.}
Given $\mathcal{R}$ the set of reference speech turns, and $\mathcal{H}$ the set of hypothesized segments, coverage is:
\begin{equation}
  \mbox{coverage}(\mathcal{R}, \mathcal{H}) = \frac{\sum_{r \in \mathcal{R}} \max_{h \in \mathcal{H}} |r \cap h|}{\sum_{r \in \mathcal{R}} |r|}
\end{equation}
where $|s|$ is the duration of segment $s$ and $r \cap h$ is the intersection of segments $r$ and $h$. Purity is the dual metric where the role of $\mathcal{R}$ and $\mathcal{H}$ are interchanged. Over-segmentation (\emph{i.e.} detecting too many speaker changes) would result in high purity but low coverage, while missing lots of speaker changes would decrease purity -- which is critical for subsequent speech turn agglomerative clustering.

\begin{figure}[htb]
  \centering
  \includegraphics[width=1.0\linewidth]{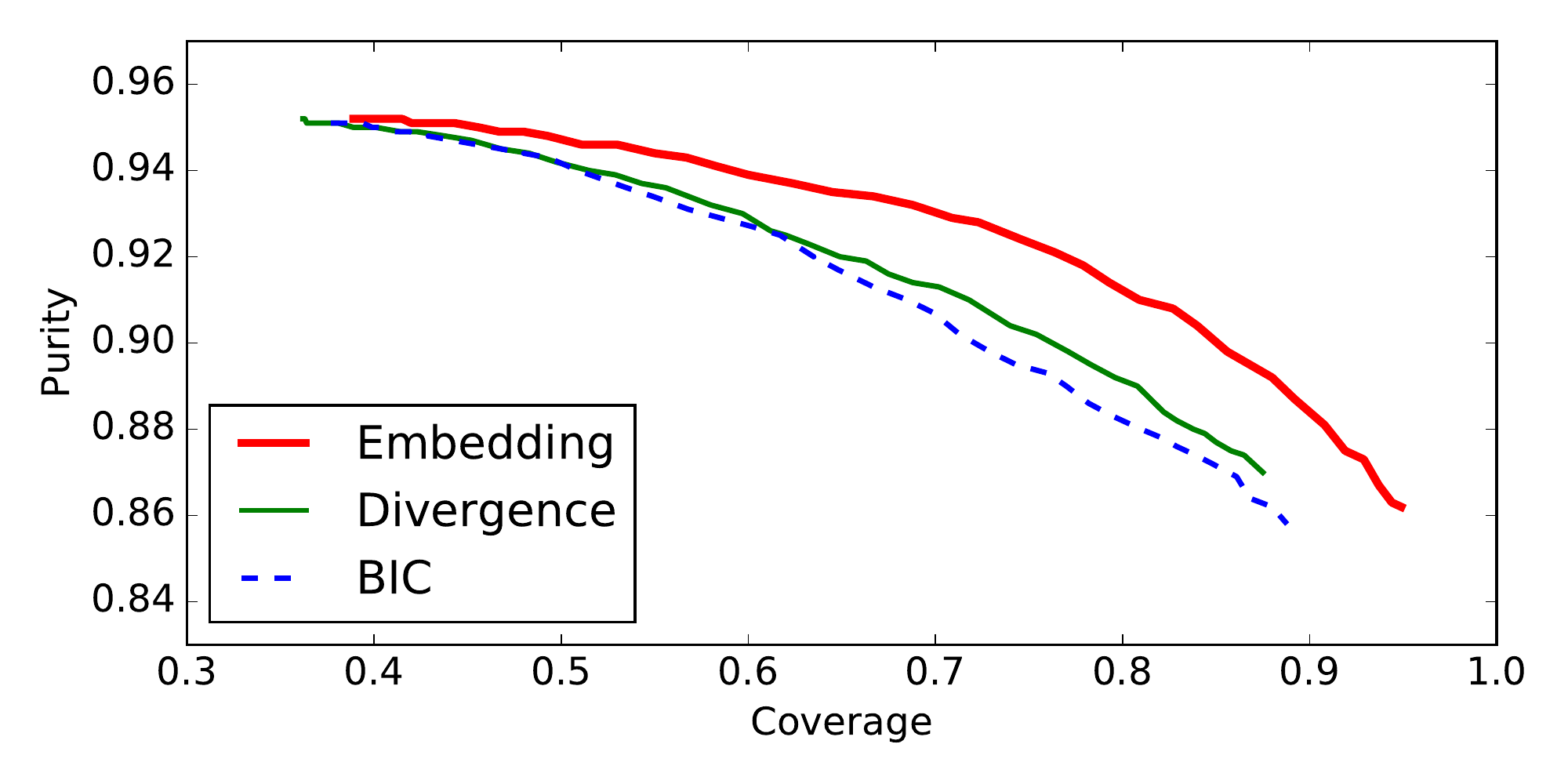}
  \caption{Performance of speaker change detection on ETAPE development set. These curves are obtained by varying the value of the final threshold.}
  \label{fig:pty_vs_cvg}
\end{figure}

\noindent\textbf{Results.}
Figure~\ref{fig:pty_vs_cvg} summarizes the results obtained when varying the value of the final threshold. Embedding-based speaker change detection clearly outperforms both BIC- and divergence-based approaches. Though it does not improve the best achievable purity (it gets $95.2\%$ vs. $95.1\%$ for divergence), embedding-based speaker change detection does improve coverage significantly. For instance, at $94.4\%$ purity, coverage is $55\%$ while BIC- and divergence-based approaches are stuck at $48\%$. In other words, it means that hypothesized speech turns are $15\%$ longer on average, with the same level of purity.

\section{Perspective}

The impact of this major improvement on the overall performance of a complete speaker diarization system (including speech activity detection and speech turn clustering) has yet to be quantified. It would also be a valuable experiment to evaluate how it generalizes to variable-length sequences (this is already supported, only not tested yet); as well as its application to speaker recognition.
Furthermore, possible future work would be to investigate the use of deeper or wider neural network architectures. Replacing the \emph{triplet loss} by the \emph{center loss} recently proposed for face recognition~\cite{Wen2016} might also be a promising research direction.

\noindent\textbf{Acknowledgement}
This work was supported by ANR through the ODESSA and MetaDaTV projects. Thanks to \emph{``LSTM guru''} Gr\'{e}gory Gelly for fruitful discussions.


\vfill\pagebreak

\bibliographystyle{IEEEbib}
\bibliography{refs}

\end{document}